\documentclass[conference, twocolumn, 10pt]{IEEEtran} 
\IEEEoverridecommandlockouts
\usepackage[utf8]{inputenc}
\usepackage{amsmath}
\usepackage{amssymb}
\usepackage{url}
\usepackage{graphicx}
\usepackage{epstopdf}
\usepackage{caption}
\usepackage{subcaption}
\usepackage{multirow}
\usepackage{amssymb}
\usepackage{setspace}
\usepackage[noadjust]{cite}
\usepackage{xspace}
\usepackage{algorithm}
\usepackage{tabularx}
\usepackage{changepage}
\usepackage{color}

\usepackage{graphicx,epsfig,color,bm}
\usepackage{color,cite}
\usepackage{amsmath,epsfig} 
\usepackage{times}
\usepackage{enumerate,type1cm}
\usepackage{amsfonts,relsize}
\usepackage{bm}
\usepackage{amssymb}
\usepackage{relsize}
\usepackage{fancybox}
\usepackage{algorithmic}
\usepackage{amssymb}
\usepackage{graphicx,epsfig,color,bm}
\usepackage{color,cite}
\usepackage{amsmath,epsfig} 
\usepackage{times}
\usepackage{enumerate,type1cm}
\usepackage{amsfonts,relsize}
\usepackage{bm}
\usepackage{amssymb}
\usepackage{relsize}
\usepackage{fancybox}
\usepackage{bbm}
\usepackage{tikz}
\usetikzlibrary{fit,positioning}
\usetikzlibrary{bayesnet}
\usetikzlibrary{shapes,decorations}
\usepackage[english]{babel}
\usepackage[xindy]{glossaries}

\def\BibTeX{{\rm B\kern-.05em{\sc i\kern-.025em b}\kern-.08em
		T\kern-.1667em\lower.7ex\hbox{E}\kern-.125emX}}

\long\def\symbolfootnote[#1]#2{\begingroup%
	\def\thefootnote{\fnsymbol{footnote}}\footnote[#1]{#2}\endgroup}

\newcommand{\beq}{\begin{equation}}
	\newcommand{\eeq}{\end{equation}}
\newcommand{\beqa}{\begin{eqnarray}}
	\newcommand{\eeqa}{\end{eqnarray}}

\newcommand{\diag}{\mathrm{diag}}

\newcommand{\Complex}{\mathbb{C}}

\usepackage{float}
\usepackage{colortbl}

	\tikzset{
		startstop/.style={
			rectangle, 
			rounded corners,
			minimum width=3cm, 
			minimum height=0.5cm,
			align=center, 
			draw=black, 
		},
		process/.style={
			rectangle, 
			minimum width=3cm, 
			minimum height=0.5cm, 
			align=center, 
			draw=black, 
		},
		decision/.style={
			rectangle, 
			minimum width=3cm, 
			minimum height=0.5cm, align=center, 
			draw=black, 
		},
		arrow/.style={thick,->,>=stealth},
		dec/.style={
			ellipse, 
			align=center, 
			draw=black, 
		},
	}
	
	
	\makeatletter
	\def\BState{\State\hskip-\ALG@thistlm}
	\makeatother
	
	\DeclareCaptionLabelSeparator{periodspace}{.\quad}
		{\end{adjustwidth}}

		\makeatletter
		\newcommand{\removelatexerror}{\let\@latex@error\@gobble}
		\makeatother
		
\newacronym{ai}{AI}{Artificial Intelligence}
\newacronym{aoa}{AOA}{angle-of-arrival}
\newacronym{aod}{AOD}{angle-of-departure}
\newacronym{ap}{AP}{access point}
\newacronym{asic}{ASIC}{application specific integrated circuit}
\newacronym{awgn}{AWGN}{additive white Gaussian noise}
\newacronym{ce}{CE}{channel estimation}
\newacronym{cw}{CW}{continuous wave}
\newacronym{cdf}{CDF}{cumulative distribution function}
\newacronym{ced}{CED}{cumulative energy density}
\newacronym{clk}{CLK}{clock}
\newacronym{csi}{CSI}{channel state information}
\newacronym{csp}{CSP}{contact service point}
\newacronym{cpu}{CPU}{central processing unit}
\newacronym{crlb}{CRLB}{Cram\'er-Rao lower bound}
\newacronym{dc}{DC}{direct current}
\newacronym{dl}{DL}{downlink}
\newacronym{dlt}{DLT}{Distributed Ledger Technology}
\newacronym{dp}{DP}{differencial privacy}
\newacronym{dram}{DRAM}{dynamic random-access memory}
\newacronym{ecc}{ECC}{elliptic curve cryptography}
\newacronym{ecdlp}{ECDLP}{elliptic curve discrete logarithm problem}
\newacronym{ecsp}{ECSP}{edge computing service point}
\newacronym{eirp}{EIRP}{equivalent isotropically radiated power}
\newacronym{em}{EM}{electromagnetic}
\newacronym{en}{EN}{energy neutral}
\newacronym{e2e}{E2E}{End-to-End}
\newacronym{fa}{FA}{federation anchor}
\newacronym{fdd}{FDD}{frequency division duplexing}
\newacronym{fhss}{FHSS}{frequency hopping spread spectrum}
\newacronym{fl}{FL}{federated learning}
\newacronym{gdpr}{GDPR}{general data protection regulation}
\newacronym{ia}{IA}{initial access}
\newacronym{ic}{IC}{integrated circuit}
\newacronym{ioe}{IoE}{Internet of Everything}
\newacronym{iot}{IoT}{Internet of Things}
\newacronym{id}{ID}{information decoding}
\newacronym{kpi}{KPI}{key performance indicator}
\newacronym{lca}{LCA}{life cycle assessment}
\newacronym{ls}{LS}{least squares}
\newacronym{lis}{LIS}{large intelligent surface}
\newacronym{liion}{Li-ion}{lithium-ion}
\newacronym{los}{LoS}{line-of-sight}
\newacronym{lmo}{LMO}{lithium ion manganese oxide}
\newacronym{mf}{MF}{matched filter}
\newacronym{ml}{ML}{machine learning}
\newacronym{mmse}{MMSE}{minimum mean square error}
\newacronym{mmwave}{mmWave}{millimeter wave}
\newacronym{mimo}{MIMO}{multiple-input multiple-output}
\newacronym{miso}{MISO}{multiple-input single-output}
\newacronym{mpc}{MPC}{multipath component}
\newacronym{mrc}{MRC}{maximum ratio combining}
\newacronym{mrt}{MRT}{maximum ratio transmission}
\newacronym{ofdm}{OFDM}{orthogonal frequency-division multiplexing}
\newacronym{ota}{OTA}{over-the-air}
\newacronym{nb}{NB}{narrowband}
\newacronym{nr}{NR}{New Radio}
\newacronym{pa}{PA}{power amplifier}
\newacronym{pdp}{PDP}{power delay profile}
\newacronym{per}{PER}{packet error rate}
\newacronym{pet}{PET}{privacy enhancing technology}
\newacronym{pki}{PKI}{public key infrastucture}
\newacronym{pl}{PL}{path loss}
\newacronym{pc}{PC}{pilot count}
\newacronym{qrrls}{QR-RLS}{QR decomposition based recursive least squares}
\newacronym{re}{RE}{radio element}
\newacronym{rf}{RF}{radio frequency}
\newacronym{rfid}{RFID}{radio frequency identification}
\newacronym{ris}{RIS}{reflective intelligent surface}
\newacronym{rms}{RMS}{root mean square}
\newacronym{rls}{RLS}{recursive least squares}
\newacronym{rss}{RSS}{received signal strength}
\newacronym{rssi}{RSSI}{received signal strength indicator}
\newacronym{rw}{RW}{RadioWeaves}
\newacronym{sa}{SA}{synchronization anchor}
\newacronym{sdg}{SDG}{Sustainable Development Goal}
\newacronym{sdn}{SDN}{software-defined network}
\newacronym{se}{SE}{spectral efficiency}
\newacronym{sinr}{SINR}{signal-to-interference-plus-noise ratio}
\newacronym{siso}{SISO}{single-input single-output}
\newacronym{slam}{SLAM}{simultaneous localization and mapping}
\newacronym{snr}{SNR}{signal-to-noise ratio}
\newacronym{srls}{SRLS}{standard recursive least squares}
\newacronym{svd}{SVD}{singular value decomposition}
\newacronym{sp}{SP}{service point}
\newacronym{sram}{SRAM}{static random-access memory}
\newacronym{tdd}{TDD}{time division duplexing}
\newacronym{tdoa}{TDOA}{time-difference-of-arrival}
\newacronym{toa}{TOA}{time-of-arrival}
\newacronym{ue}{UE}{user equipment}
\newacronym{uhf}{UHF}{ultra high frequency}
\newacronym{ula}{ULA}{uniform linear array}
\newacronym{upa}{UPA}{uniform planar array}
\newacronym{ul}{UL}{uplink}
\newacronym{ura}{URA}{uniform rectangular array}
\newacronym{uwb}{UWB}{ultrawideband}
\newacronym{wb}{WB}{wideband}
\newacronym{wpt}{WPT}{wireless power transfer}
\newacronym{zf}{ZF}{zero forcing}

\newacronym{lpt}{LPT}{laser power transfer}
\newacronym{rfpt}{RFPT}{radio frequency power transfer}
\newacronym{ipt}{IPT}{inductive power transfer}
\newacronym{cpt}{CPT}{capacitive power transfer}
\newacronym{apt}{APT}{acoustic power transfer}
\newacronym{cfo}{CFO}{carrier frequency offset}
\newacronym{lo}{LO}{local oscillator}
\newacronym{if}{IF}{intermediate frequency}
\newacronym{duc}{DUC}{digital up conversion}
\newacronym{ddc}{DDC}{digital down conversion}
\newacronym{dsp}{DSP}{digital signal processing}
\newacronym{vco}{VCO}{voltage-controlled oscillator}
\newacronym{pll}{PLL}{phase-locked loop}
\newacronym{lna}{LNA}{low noise amplifier}
\newacronym{dac}{DAC}{digital-to-analog converter}
\newacronym{adc}{ADC}{analog-to-digital converter}
\newacronym{de}{DE}{drain efficiency}
\newacronym{pae}{PAE}{power-added efficiency}
\newacronym{sar}{SAR}{specific absorption rate}
\newacronym{mppt}{MPPT}{maximum power point tracking}
\newacronym{isi}{ISI}{intersymbol interference}
\newacronym{pps}{PPS}{pulse per second}
\newacronym{icnirp}{ICNIRP}{International Commission on Non-Ionizing Radiation Protection}
\newacronym{fd}{FD}{Front-haul distance}
\newacronym{wd}{WD}{Wireless distance}
\newacronym{ntp}{NTP}{Network Time Protocol}
\newacronym{ptp}{PTP}{Precision Time Protocol}
\newacronym{wr}{WR}{White Rabbit}

\title{Data-Driven Robust Beamforming for Initial Access
\thanks{This work was funded by the REINDEER project of the European Union's Horizon 2020 research and innovation program under grant agreement No. 101013425.}
}
\author{\IEEEauthorblockN{Sai Subramanyam Thoota}
	\IEEEauthorblockA{\textit{Dept. of Electrical Engineering (ISY)} \\
		\textit{Link\"oping University}\\
		581 83 Link\"oping, Sweden \\
		sai.subramanyam.thoota@liu.se}
	\and
	\IEEEauthorblockN{Joao Vieira}
	\IEEEauthorblockA{\textit{Ericsson Research} \\
		223 62 Lund, Sweden \\
		joao.vieira@ericsson.com}
	\and
	\IEEEauthorblockN{Erik G. Larsson}
	\IEEEauthorblockA{\textit{Dept. of Electrical Engineering (ISY)} \\
		\textit{Link\"oping University}\\
		581 83 Link\"oping, Sweden \\
		erik.g.larsson@liu.se}
}

\begin{document}

\maketitle
\date{}
\begin{abstract}
	We consider a robust beamforming problem where large amount of \gls{dl} \gls{csi} data available at a multiple antenna \gls{ap} is used to improve the link quality to a \gls{ue} for beyond-5G and 6G applications such as environment-specific \gls{ia} or \gls{wpt}. As the \gls{dl} \gls{csi} available at the current instant may be imperfect or outdated, we propose a novel scheme which utilizes the (unknown) correlation between the antenna domain and physical domain to localize the possible future \gls{ue} positions from the historical \gls{csi} database. Then, we develop a codebook design procedure to maximize the minimum sum beamforming gain to that localized \gls{csi} neighborhood. We also incorporate a \gls{ue} specific parameter to enlarge the neighborhood to robustify the link further. We adopt an indoor channel model to demonstrate the performance of our solution, and benchmark against a usually optimal (but now sub-optimal due to outdated \gls{csi}) \gls{mrt} and a subspace based method. We numerically show that our algorithm outperforms the other methods by a large margin. This shows that customized environment-specific solutions are important to solve many future wireless applications, and we have paved the way to develop further data-driven approaches. 
\end{abstract}
\begin{IEEEkeywords}
6G, beyond-5G, codebook, data-driven, initial access, robust beamforming.
\end{IEEEkeywords}

\section{Introduction} \label{sec:Introduction}

A fundamental problem with \acrfull{dl} beamforming in a \gls{mimo} system is that one may not have access to highly accurate \acrfull{csi} at the transmitter, e.g. at a multi-antenna \acrfull{ap} or transmit/receive point of a distributed \gls{mimo} system. For example, in an environment with mobility, the \gls{csi} estimate may have been obtained some time ago, and beamforming with such an outdated \gls{csi} will reduce the beamforming gain. In particular, if the phase of the channel estimate is outdated, then the beamforming gain may be lost completely. 

Beamforming with outdated \gls{csi} is a crucial problem in several next generation wireless applications such as {environment-specific \acrfull{ia}}, \acrfull{wpt} from infrastructure to passive devices, backscattering communication with zero-energy devices via multi-antenna emitters (and readers), and in {\gls{dl} \gls{mimo} communications where the time interval between the channel estimation and \gls{dl} precoder computation is comparable to the coherence time of the channel~\cite{Ahmet_Globecom_2022,Perera_IEEECOMTUTs_2018}}. Beam management during \gls{ia} is another important topic in 5G-NR and study items on using data-driven approaches for the same has been a focus area in release $18$ and later. Also, codebook-based beam sweeping procedures for \gls{ia} are employed to lock the best pair of beams between the \glspl{ap} and the \glspl{ue} in 5G-NR, and using data-driven methods to design application specific codebooks will play an increasingly important role in beyond-5G and 6G applications~\cite{Giordani_COMMSURVs_2019,3GPP_TR38843_2017NR,3GPP_TR38843_2022NR}.

Typical approaches to tackle this outdated \gls{csi} problem are to use diversity transmission techniques such as space-time codes or subspace tracking methods using covariance matrices such as eigen-beamforming. However, many real world channels are not well modeled by second order statistics, and hence parameterization using a linear subspace may not be an appropriate approach. 

%

In this paper, we focus on data-driven robust beamforming in an environment when one or more \glspl{ap} are deployed at distinct geographical locations with mobile \glspl{ue}. In many examples, these \glspl{ue} may be moving across similar but not identical trajectories. One primary deployment case would be indoors, for example machines in a factory, or forklifts that move around in a warehouse. Robust beamforming is of concern during several phases of the communication, including \gls{ia}, data transmission, and \gls{wpt} and communication with passive devices (where a power beam needs to be beamformed to a device whose channel is only approximately known). 

We consider a scenario where a database of historical \gls{dl} \gls{csi} is available at the \glspl{ap}. We obtain this database by direct \gls{ul} channel measurements and/or \gls{dl} channel measurements plus \gls{ue}-feedback. The time instants at which the \gls{csi} are recorded may even be sporadic based on the energy availability or the wake up/sleep cycles of the \glspl{ue}. Therefore, when the \gls{ap} gets a \gls{dl} transmission trigger, the most recent \gls{csi} estimate from the database will be outdated due to which an usually optimal \acrfull{mrt} beamformer will be sub-optimal. However, we can potentially utilize the knowledge of the repeated \gls{ue} trajectory to design beamforming vectors.

We propose a solution which makes use of the knowledge of the historical \gls{csi} to improve the link quality (beamforming gain) to a set of candidate locations estimated where the \gls{ue} may be. If the physical location of the \gls{ue} is known, then we can predict the future locations using its mobility pattern. However, we do not have knowledge of the position of the \gls{ue}. Therefore, we proceed by deducing the possible future locations using an unknown mapping from the \gls{csi} space to the physical domain. Note that our solution does not depend on any sensing or positioning technique. However, our solution can be enhanced by using any such side information. 

The primary idea is to design a codebook of beamforming {(or precoding)} vectors, to robustly transmit information and/or power to such set of candidate locations. One of our novel contributions is that, from the historical \gls{csi}, we form a database of channel responses seen in the past, preferably in chronological order (each response is typically associated with a particular location of the user and a particular time-stamp). Then, after a request for \gls{dl} transmission is triggered, the most recently measured channel estimate (which might be slightly outdated) is fetched from the database, as well as other previously seen channel estimates that are ``similar'' to the most recently measured channel estimate. With that, a set of responses that are close in a specific distance metric to the most recent channel estimate, as well as to the other ``similar'' channel estimates, is formed. Based on this so-obtained set of responses, a codebook of beamforming vectors is determined through the use of an optimization algorithm. This set of beamformers are designed according to a minimax principle such that they are good for the ``worst'' among the set of channel responses. 

Note that the only inputs that are mandatory for the proposed method are the \gls{csi} estimates (which are necessary to form the \gls{csi} database). Optional inputs to the proposed method are the timestamps associated to \gls{csi} estimates. Though we demonstrate the solution with a single \gls{ap}, the solution is equally applicable in any multi-antenna setup with one or multiple \glspl{ap}, either in a co-located or distributed \gls{mimo} deployment with appropriate modifications. In the remainder of the paper, we only use the term \gls{ap} for simplicity. 

{A codebook design problem to improve the coverage probability with geometric channel models has been studied in \cite{Mustafa_VTCS_2022} where the authors have designed heuristic algorithms to design beamformers. However, they have not used a database explicitly to build a channel subset in order to design a beamforming codebook. To the best of our knowledge, a principled approach to design a data-driven beamforming codebook based on an environment or application specific problem has not been studied in the literature yet. Our data-driven robust beamforming solution provides a good benchmark and motivation to design more data-driven approaches for several beyond-5G and 6G wireless applications}. {We also mention that the aspects such as the overhead associated with the channel estimation and the \gls{csi} database generation need a separate study and are beyond the scope of this paper.}

\section{Data-Driven Robust Beamforming System} \label{sec:SysModel}
\begin{figure}
	\centering
	\includegraphics[width=\columnwidth]{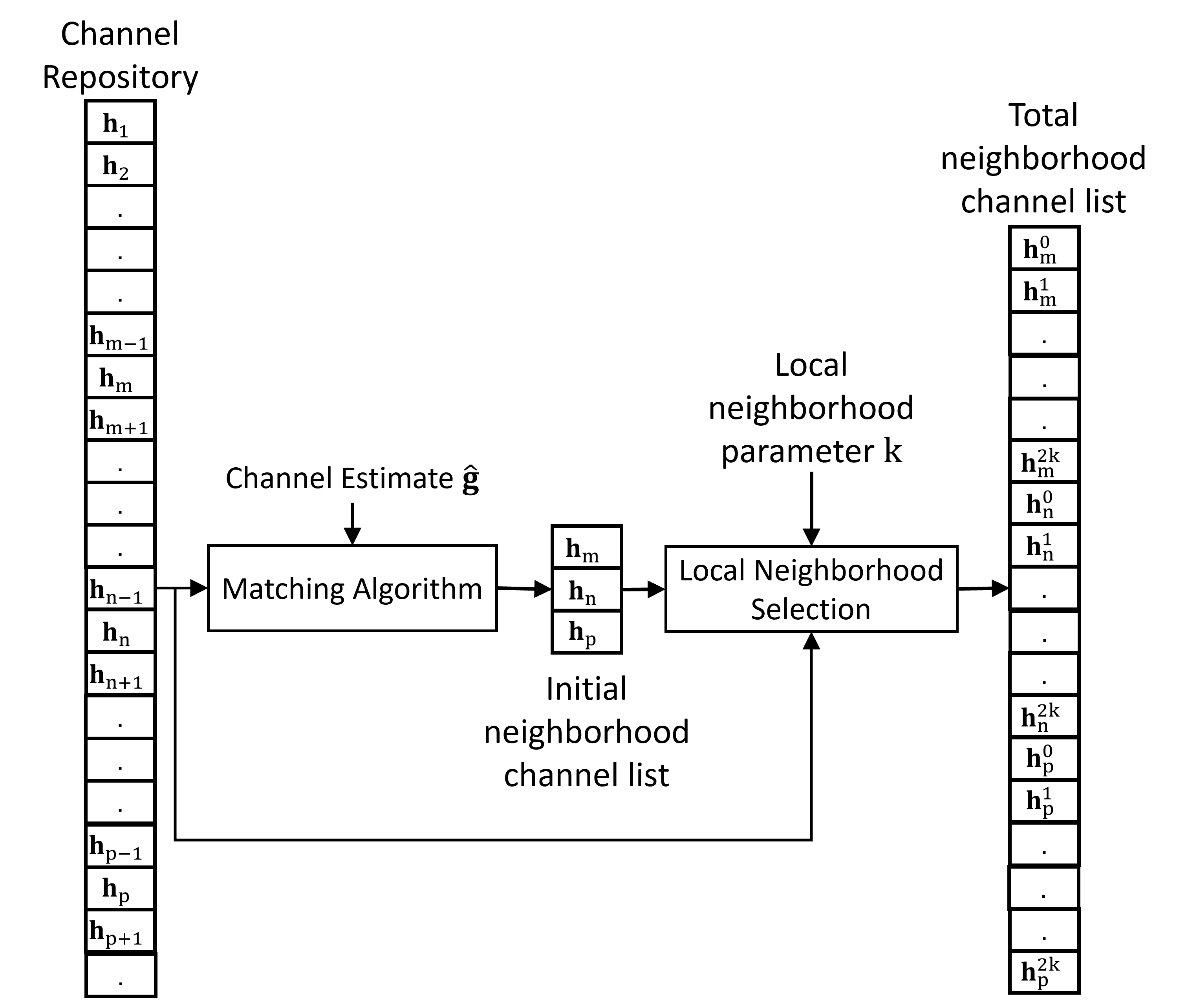}
	\caption{Neighborhood channel list generation procedure.}
	\label{fig:NeighborhoodSelection}
	\vspace{-0.25cm}
\end{figure}
We consider a wireless communication system where one $M$ transmit antenna \gls{ap} beamforms to a single receive antenna \gls{ue} for data or power transfer. We envisage a situation where the \gls{ue} traverses in a trajectory repeatedly with minor perturbations which may occur due to some mechanical imperfections. The \gls{ap} records the \gls{ue}'s \gls{csi} in a database chronologically using any channel estimation mechanism. For instance, in a \gls{tdd} based system with channel reciprocity, the \gls{ap} can estimate the \gls{dl} \gls{csi} using the pilots transmitted by the \gls{ue}. Till the \gls{ap} gets a trigger for \gls{dl} transmission, it estimates and records the \gls{csi} in the database continuously. Once it gets a downlink transmission trigger, 
the \gls{ap} gets the latest \gls{csi} from the database. Then, it runs a neighborhood channel generation algorithm which outputs a list of channel vectors from the database. As we know apriori that the \gls{ue} has followed a trajectory, the neighborhood channel list obtained corresponds to the set of channel vectors which are potential candidates that capture the long-term characteristics of a local neighborhood of the current outdated \gls{csi}. Once the neighborhood list is generated, the codebook of beamforming vectors is designed, and then the downlink operation is executed.

We illustrate the neighborhood channel list generation procedure mentioned above in the Figure~\ref{fig:NeighborhoodSelection}. First, we define a closeness-metric between the current outdated \gls{csi} $\widehat{\mathbf{g}}$ and the $i^\text{th}$ element in the channel repository $\mathbf{h}_i$ as 
\begin{align}
	c_i = \frac{\left\arrowvert\widehat{\mathbf{g}}^H\mathbf{h}_i\right\arrowvert^2}{\left\Arrowvert\widehat{\mathbf{g}}\right\Arrowvert^2 \left\Arrowvert\mathbf{h}_i\right\Arrowvert^2}.\label{eqn:closenessMetric}
\end{align}

We input the channel repository (in the left side of the Figure~\ref{fig:NeighborhoodSelection}) to a matching algorithm which chooses an initial neighborhood channel list given the outdated \gls{csi} $\widehat{\mathbf{g}}$. The matching algorithm computes the closeness-metric between $\widehat{\mathbf{g}}$ and each element of the channel repository and chooses all the elements whose closeness-metrics are greater than a predefined threshold. For example, in Figure~\ref{fig:NeighborhoodSelection}, $\{m,n,p\}$ are the indices of the selected entries from the database to form the initial neighborhood channel list. We show only three selected indices for illustration purposes, but the matching algorithm can choose any number of vectors in the initial neighborhood channel list. 

To increase the robustness, we input these selected indices and the associated channel repository to a local neighborhood selection algorithm. We also input a predefined parameter $k$ to the local neighborhood selection process.~\footnote{To illustrate our data-driven robust beamforming solution, we fix the parameter $k$ in our paper. However, it can also be learnt dynamically based on the \gls{ue}'s mobility statistics.} This selection algorithm executes as many times as the number of entries in the initial neighborhood channel list. For instance, in the first run, it picks the first index from the list (that is $m$) and goes to that index in the channel repository. Then it picks $k$ channels each before and after the $m$-th entry, and places each one in a new neighborhood list. Therefore, for each element from the initial neighborhood list, it picks $2k$ elements from the channel repository in total. This is illustrated in the right side of the figure. Here, the subscripts indicate the elements in the initial neighborhood list and the superscripts represent the local neighborhood. For example, $\{\mathbf{h}_m^0,\mathbf{h}_n^0,\mathbf{h}_p^0\}$ are the same as the initial neighborhood list $\{\mathbf{h}_m,\mathbf{h}_n,\mathbf{h}_p\}$, whereas $\{\mathbf{h}_m^1,\ldots,\mathbf{h}_m^{2k}\}$ are the elements $\{\mathbf{h}_{m-k},\mathbf{h}_{m-k+1},\ldots,\mathbf{h}_{m-1},\mathbf{h}_{m+1},...,\mathbf{h}_{m+k-1},\mathbf{h}_{m+k}\}$ from the channel repository. 

The neighborhood selection algorithm repeats this local neighborhood selection process for all the other elements in the initial neighborhood channel list. Finally, the output of this step is a total neighborhood channel list which is at most of size $6k+3$ $(=3(2k+1))$ in the illustrated example. In general, if the number of elements in the initial neighborhood list is $T$, then the maximum length of the total neighborhood channel list is $(2k+1)T$. Since there can be overlaps between the local neighborhoods associated with different elements in the initial list, the length of the total neighborhood list can be smaller than $(2k+1)T$.

Once we obtain the total neighborhood channel list, our next step is to design a codebook of beamforming vectors to maximize a chosen objective function. We describe the codebook design next. Let us define the number of vectors output by the neighborhood list generation process by $K$. Without loss of generality, we denote the channel vectors in the final neighborhood list by $\{\mathbf{h}_1, \ldots, \mathbf{h}_K\}$ in the subsequent sections of this paper. Note that this neighborhood list generation procedure does not take into account any anomalies in the \gls{ue}'s movements. However, we can incorporate such changes to refine the neighborhood by periodically learning the \gls{ue}'s mobility behavior which is part of our future work.

\section{Robust Beamforming Codebook Design}\label{sec:ProposedApproach}

In this section, our goal is to design a beamforming codebook to provide a robust link between the \gls{ap} and the \gls{ue}. We define the link robustness using the minimum of the sum beamforming gain across all the elements in the \gls{csi} neighborhood list. Note that our proposed data-driven robust beamforming procedure is equally applicable to any other optimization criterion. If $\widehat{\mathbf{g}}$ is the exact \gls{csi}, then \gls{mrt} is the optimal transmission scheme. However, as mentioned before, in general $\widehat{\mathbf{g}}$ differs from the true channel, and the idea is that the channel responses in the neighborhood list are close to the true channel as the \gls{ue} will likely be at a location close to some location that it has visited before, for which \gls{csi} is available in the neighborhood list. Therefore, we design a codebook of beamforming vectors which yields robustness of the beamforming process. We can then choose to beamform either with one or many vectors from the codebook. Let us define the number of vectors in the codebook by $L$.

Mathematically, we represent the robust beamforming codebook problem as
\begin{align}
	\mathcal{P}_0: \qquad&\max_{\{\mathbf{f}_1,\ldots,\mathbf{f}_L\}\in\mathbb{C}^{M \times L}} \min_{i\in [K]} \frac{\sum_{\ell=1}^L\arrowvert\mathbf{h}_i^H \mathbf{f}_\ell\arrowvert^2}{\Arrowvert \mathbf{h}_i\Arrowvert^2}\label{eqn:maxminProb1}\\
	&\text{s. t. }\quad\Arrowvert\mathbf{f}_\ell\Arrowvert^2\le P,\quad\ell\in[L],\nonumber
\end{align}
where $\{\mathbf{f}_1, \ldots, \mathbf{f}_L\}$ are the beamforming vectors to be designed, and $P$ is the transmit power constraint at the \gls{ap} per channel use.

{Intuitively, this optimization metric is a quantitative measure of how much energy the \gls{ue} accumulates over $L$ channel uses. For instance, a passive device stores the energy received from the infrastructure in a battery during the \gls{ia} phase and transmits when sufficient energy is available. We design the beamformers to maximize the accumulated energy by the worst candidate channel within the neighborhood of the current outdated \gls{csi}. This approach of maximizing the minimum sum beamforming gain provides the link robustness. If \gls{ue} does not have the capability to accumulate the received energy (for example, a backscatter device), then a different optimization metric can be chosen accordingly. However, the neighborhood channel list generation procedure is equally applicable to any optimization criterion.}

Problem $\mathcal{P}_0$ is a non-convex max-min-sum optimization problem which does not have any closed form analytical solution. Hence, we need to employ convex relaxation or bounding techniques to obtain a locally optimal solution. 

To do that, we first write an equivalent epigraph form of $\mathcal{P}_0$ as
\begin{align}
	\mathcal{P}_1: \qquad&\max_{t,\, \{\mathbf{f}_1,\ldots,\mathbf{f}_L\}\in\mathbb{C}^{M \times L}}\, t\label{eqn:maxMinProb2}\\ &\text{s. t. } \frac{\sum_{\ell=1}^L\arrowvert\mathbf{h}_i^H \mathbf{f}_\ell\arrowvert^2}{\Arrowvert \mathbf{h}_i\Arrowvert^2}\ge t,\quad i\in[K],\label{eqn:probP1_1}\\&\qquad\Arrowvert\mathbf{f}_\ell\Arrowvert^2\le P, \quad \ell\in[L].\nonumber
\end{align}

One of the approaches to solve $\mathcal{P}_1$ is to solve its inverse problem iteratively, which is to minimize the transmit power subject to a minimum beamforming gain target~\cite{Karipidis_TSP_2008,Christopoulos_SPAWC_2015,MinDong_TSP_2020}. We state the inverse quality-of-service (QoS) problem to $\mathcal{P}_1$:
\begin{align}
	\mathcal{D}_1:\quad &\min_{t,\mathbf{f}_1,\ldots,\mathbf{f}_L} \sum_{\ell=1}^L \Arrowvert\mathbf{f}_\ell\Arrowvert^2\label{eqn:minPowerDualProb1}\\ &\text{s. t. } \frac{\sum_{\ell=1}^L\arrowvert\mathbf{h}_k^H \mathbf{f}_\ell\arrowvert^2}{\Arrowvert \mathbf{h}_k\Arrowvert^2}\ge t,\quad k\in[K].\nonumber
\end{align}


An optimal beamforming structure to solve $\mathcal{D}_1$ has been solved in \cite{MinDong_TSP_2020}. The solution which is a special case of multicast beamforming takes the form of a weighted minimum mean squared error filter structure. 
It modifies the dimension of the optimization problem from the number of transmit antennas to the size of the channel dataset $K$. 
%
However, in a data-driven robust beamforming problem, the size of the dataset can be much larger than the number of antennas due to which this transformation may increase the complexity of the problem, and therefore choosing this approach is not appropriate in this context. 
Hence, we solve the original problem $\mathcal{P}_1$ directly.

As mentioned earlier, $\mathcal{P}_1$ is non-convex in its original form, and therefore we convexify the constraint \eqref{eqn:probP1_1} and solve the resulting problem in an iterative manner till a suitable convergence criterion is satisfied. To convexify the non-convex constraint, we first choose an initial feasible point $\{\mathbf{w}_1,\ldots,\mathbf{w}_L\}\in\Complex^{M\times L}$ and apply a first order Taylor series approximation of $\sum_{\ell=1}^L\arrowvert\mathbf{h}_i^H \mathbf{f}_\ell\arrowvert^2$, $\forall \ell$, around it. The resulting optimization problem is:
\begin{align}
	\mathcal{P}_2: \qquad&\max_{t,\, \{\mathbf{f}_1,\ldots,\mathbf{f}_L\}\in\mathbb{C}^{M \times L}} \,t\label{eqn:maxMinProb3}\\ 
	&\text{s. t. } \sum_{\ell=1}^L\left(\mathbf{w}_\ell^H\mathbf{h}_k\mathbf{h}_k^H\mathbf{w}_\ell - 2\Re\left(\mathbf{f}_\ell^H\mathbf{h}_k\mathbf{h}_k^H\mathbf{w}_\ell\right)\right)\nonumber\\&\qquad\qquad\qquad\qquad\qquad\quad\le -t\Arrowvert \mathbf{h}_k\Arrowvert^2,\quad k\in[K],\nonumber\\&\qquad\Arrowvert\mathbf{f}_\ell\Arrowvert^2\le P, \quad \ell\in[L].\nonumber
\end{align}
where $\Re(x)$ denotes the real part of a complex scalar $x$. 

{This approach of linearizing using a first-order Taylor series expansion provides a lower-bound (or a minorizing function) to the sum of quadratic functions (whose minimum we want to maximize) within the feasible region. As the minimum of the sum of multiple linear functions is a concave function, it can be maximized using convex optimization procedures.} We solve the optimization problem $\mathcal{P}_2$ using the convex optimization solver ``cvx'' to obtain a locally optimal solution $\{\mathbf{f}_1^{(o)},\ldots,\mathbf{f}_L^{(o)}\}$~\cite{cvxr,gb08}. We now substitute this solution as the new $\{\mathbf{w}_1,\ldots,\mathbf{w}_L\}$, and solve it iteratively till a suitable convergence criterion is satisfied. We can show that this approach converges to a stationary point of the original non-convex optimization problem $\mathcal{P}_0$~\cite{marks1978general,Hunter_AmerStat_2004}. We summarize the data-driven robust beamforming in Algorithm~\ref{Alg:DDRBF}.

\begin{algorithm}[!t]
	\caption{Data-Driven Robust Beamforming} \label{Alg:DDRBF}
	\begin{algorithmic}[1]
		\renewcommand{\algorithmicrequire}{\textbf{Input:}}
		\renewcommand{\algorithmicensure}{\textbf{Output:}}
		\REQUIRE $\mathbf{h}_1,\ldots,\mathbf{h}_K$, $P$.
		\ENSURE  $\mathbf{f}_1^{(o)},\ldots,\mathbf{f}_L^{(o)}$.
		\STATE \textbf{Initialize} $\mathbf{w}_1,\ldots,\mathbf{w}_L$ which satisfies the transmit power constraint $P$.
		\REPEAT
		\STATE Solve $\mathcal{P}_2$ using cvx to obtain $\mathbf{f}_1^{(o)},\ldots,\mathbf{f}_L^{(o)}$.
		\STATE $\mathbf{w}_1 = \mathbf{f}_1^{(o)},\ldots,\mathbf{w}_L = \mathbf{f}_L^{(o)}$.
		\UNTIL stopping condition is met	
	\end{algorithmic}
\end{algorithm}

\section{Simulation Results} \label{sec:SimResults}
\begin{figure}[!t]
	\centering
	\includegraphics[width=\columnwidth]{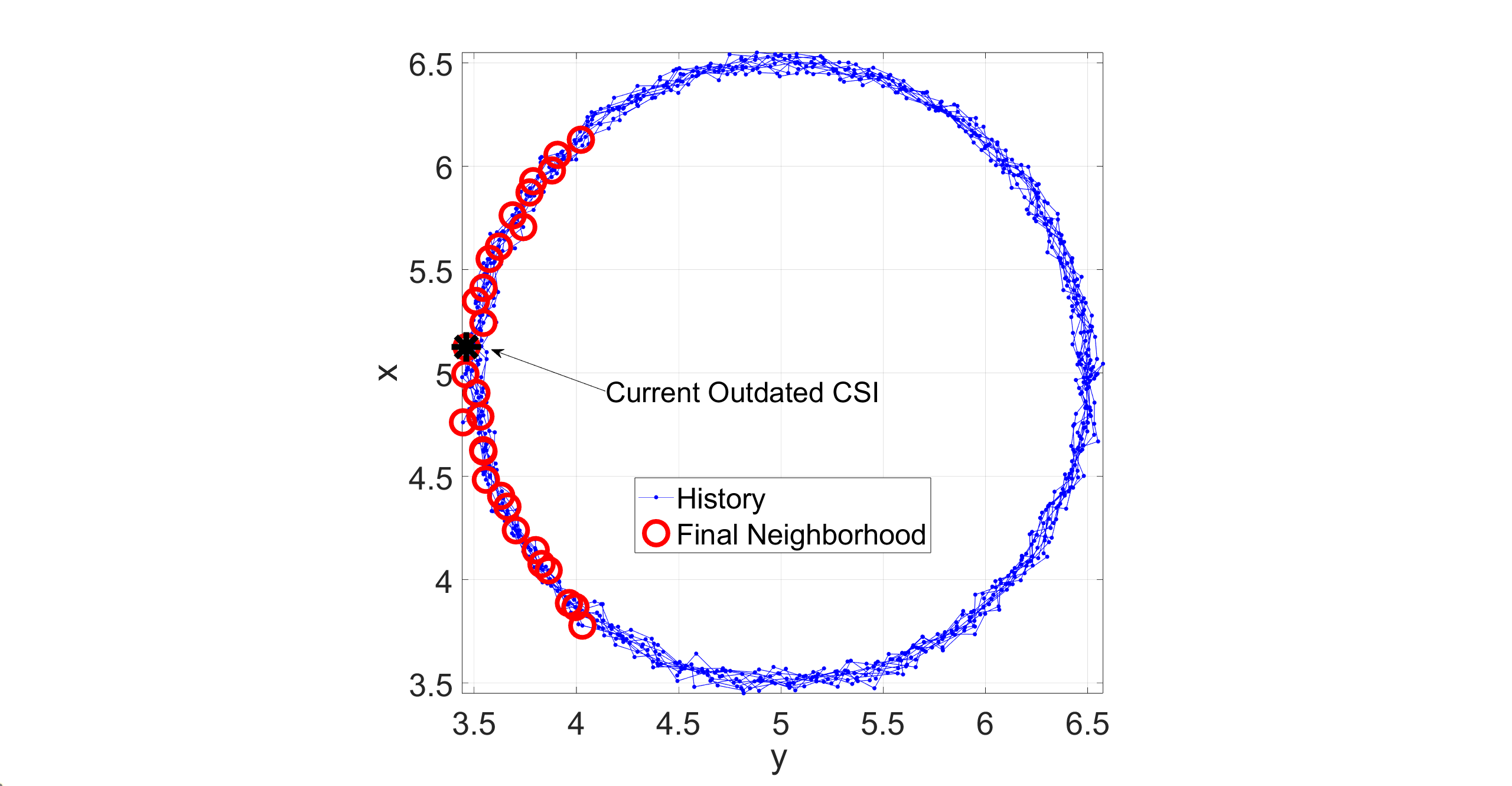}
	\caption{Example of neighborhood channel list generation for a UE moving in a circular trajectory.}
	\label{fig:CircPath_5globalNbrs_20localnbrs_1000points}
	\vspace{-0.2cm}
\end{figure}
\begin{figure}[!t]
	\centering
	\includegraphics[width=\columnwidth]{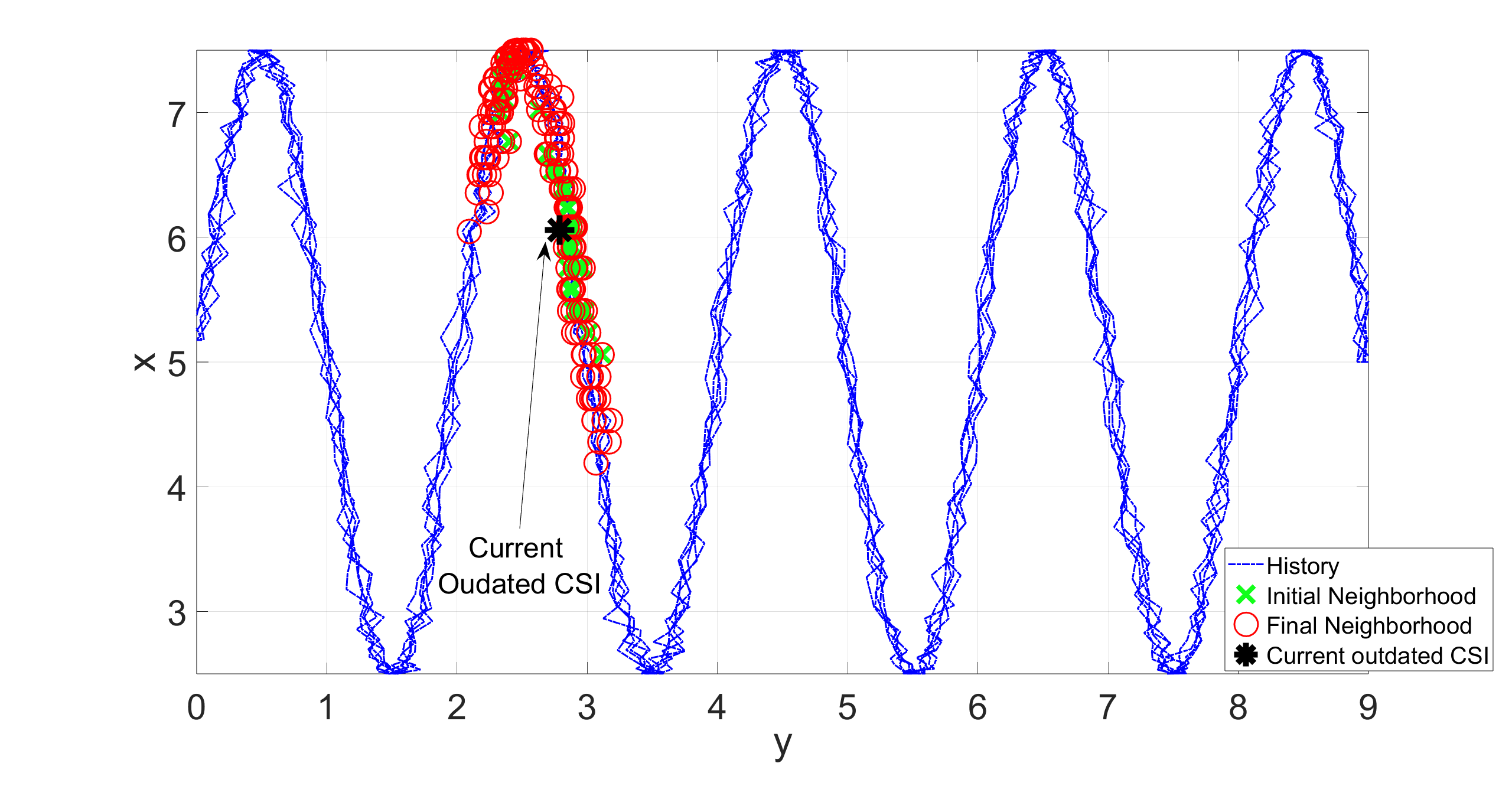}
	\caption{Example of neighborhood channel list generation for a UE moving in a zig-zag trajectory.}
	\label{fig:SinePath_AngleThreshold50deg_ClosenessMetricPoint6428_10localnbrs_2000points}
	\vspace{-0.35cm}
\end{figure}

We describe the simulation scenario and the narrowband frequency-flat channel model used to demonstrate the data-driven robust beamforming solution. We adopt a channel model that captures the typical indoor radio environments through deterministic, specular multipath components (SMCs) and stochastic, diffuse/dense multipath components (DMCs)~\cite{Wilding_Sensors_2018,Deutschmann_ICCWS_2022}. We consider a large room of dimensions $5$m $\times$ $9$m $\times$ $3.5$m with a \gls{ura} placed on a wall centered at $(5\text{m},0\text{m},1\text{m})$. The width and height of the \gls{ura} are $2.5$~m and $1.5$~m, respectively. The carrier frequency is set to $2.4$~GHz and the antennas are spaced half a wavelength apart from each other.

\subsection{Indoor Channel Model}
We employ a memoryless \gls{miso} channel model between the \gls{ap} and the \gls{ue} as
\begin{align}
	\mathbf{h} = \sum_{s=1}^S \mathbf{h}_s + \sum_{s=1}^S \mathbf{h}_{sc,s},\label{eqn:ChannelModel1}
\end{align}
where the first and second summations correspond to the SMCs and DMCs, respectively, and $S$ is the number of virtual image sources including the original \gls{ura}. The \gls{ura} is mirrored at the walls to obtain the virtual image sources~\cite{Leitinger_JSAC_2015}. The deterministic SMCs are modeled based on an image source model combined with an environment floor plan. The $m$-th element of the $s$-th SMC component is given by
\begin{align}
	\left[\mathbf{h}_s\right]_m = \frac{\lambda}{4\pi d_{s,m}} g_{SMC,s} \exp\left(-j\frac{2\pi}{\lambda}d_{s,m}\right),
\end{align}
where $d_{s,m}$ is the distance between the transmit antenna $m\in\{1,\ldots,M\}$ of the $s$-th image source and the \gls{ue}, $\lambda$ is the wavelength in meters, $g_{SMC,s}$ is the complex gain associated with reflection $s$, and the exponential term represents the phase shift due to the propagation distance. In simulations, we have set $g_{SMC,s}$ to $-3$~dB.

The DMCs model the stochastic scattering effects and are modeled by a random number of $N_{sc}$ point scatterers which are Poisson distributed with mean $10$~m$^{-3}$ in our simulations. A single ellipsoid is positioned at ($5$m, $8.75$m, $1$m) with semi-axes ($1.5$m, $0.5$m, $1.5$m) and the scatterers are placed on it to mimick a rough surface. We assume only single-bounce scattering similar to that mentioned in \cite{Deutschmann_ICCWS_2022}. The DMC component for the $s$-th SMC is defined as
\begin{align}
	\mathbf{h}_{sc,s} = \mathbf{H}_{TX,s}^T \boldsymbol{\Sigma}_{sc} \mathbf{h}_{RX},
\end{align}
where $\mathbf{H}_{TX,s}\in\mathbb{C}^{N_{sc}\times M}$ and $\mathbf{h}_{RX}\in\Complex^{N_{sc}\times 1}$ are the channels from the $s$-th image array to the scatterers and scatterers to the \gls{ue}, respectively. The diagonal matrix $\boldsymbol{\Sigma}_{sc}=\diag\left(\sqrt{\sigma_1}\exp\left(j\varphi_1\right),\ldots,\sqrt{\sigma_{N_{sc}}}\exp\left(j\varphi_{N_{sc}}\right)\right)\in\Complex^{N_{sc}\times N_{sc}}$ contains the log-normally distributed radar cross sections $\{\sigma_1,\ldots,\sigma_{N_{sc}}\}$ with mean $\mu_{sc}$ and variance $\sigma_{sc}^2$, and the i.i.d uniformly distributed phase shifts (between $0$ and $2\pi$) $\{\varphi_1,\ldots,\varphi_{N_{sc}}\}$ of the point scatterers. The values for $\mu_{sc}$ and variance $\sigma_{sc}^2$ are set to $10^2\pi$~cm$^2$ and $20\pi$~cm$^2$, respectively. The $(\ell,m)$-th and the $m$-th entries of $\mathbf{H}_{TX,s}$ and $\mathbf{h}_{RX}$ are:
\begin{align}
	\left[\mathbf{H}_{TX,s}\right]_{\ell,m} &= \frac{1}{\sqrt{4\pi} d_{s,\ell,m}} g_{SMC,s} \exp\left(-j\frac{2\pi}{\lambda}d_{s,\ell,m}\right),\\
	\left[\mathbf{h}_{RX}\right]_m &= \frac{\lambda}{4\pi d_{m}}  \exp\left(-j\frac{2\pi}{\lambda}d_{m}\right),
\end{align}
respectively, where $d_{s,\ell,m}$ is the distance between the $\ell$-th scatterer and the $m$-th antenna of the $s$-th image source, and $d_m$ is the distance between the scatterer and the \gls{ue}. 

\subsection{Examples of Neighborhood Channel List Generation}
Now, we give two examples to demonstrate the output of the neighborhood channel list generation process described in Section~\ref{sec:SysModel}. We generate the current outdated \gls{csi} $\widehat{\mathbf{g}}$ randomly in a similar trajectory as the channel database. Figure~\ref{fig:CircPath_5globalNbrs_20localnbrs_1000points} shows the \gls{ue}'s positions traversing in a circular path on the ground ($z=0$). The units in the $x$ and $y$ axes are in meters. The black colored marker represents the current outdated \gls{csi} which is input to the neighborhood list generation process. We generate the outdated \gls{csi} randomly in the same noisy path traversed by the \gls{ue}. We set the lengths of the initial neighborhood list and the local neighborhood parameter to $5$ and $5$, respectively, to demonstrate the output of the neighborhood list generation process. The red colored circles are the output of the neighborhood list generation procedure. This exemplifies that there is a mapping between the antenna and physical domains which can be exploited to predict the \gls{ue}'s future positions in a reliable way. Figure~\ref{fig:SinePath_AngleThreshold50deg_ClosenessMetricPoint6428_10localnbrs_2000points} shows the \gls{ue}'s trajectory when it moves around the room in a zig-zag fashion. We also show the output of the initial neighborhood generation procedure (green colored markers) to show the effectiveness of exploiting the correlation between the current outdated \gls{csi} to predict the possible future locations of the \gls{ue}. {Note that the final neigborhood list may not be symmetrically distributed around the current outdated \gls{csi} due to the stochastic behavior of the scatterers and the \gls{ue}'s trajectory}.

\subsection{Numerical Results for Robust Beamforming}
We benchmark the performance of the max-min-sum (MMS) beamforming with eigen-beamforming (EBF) and \gls{mrt}. We set the maximum transmit power to $0$~dBW per channel use. For the \gls{mrt} scheme, we project the current outdated \gls{csi} on to each element in the neighborhood channel list and pick the least gain, and then multiply it with the size of the codebook. We choose the least gain since it captures the worst position of the \gls{ue} if it is beamformed with the current outdated \gls{csi}. For the EBF, we compute the covariance matrix of the channels in the neighborhood list and pick the $L$ dominant eigenvectors as the beamforming codebook.

\begin{figure}[!t]
	\centering
	\includegraphics[width=\columnwidth]{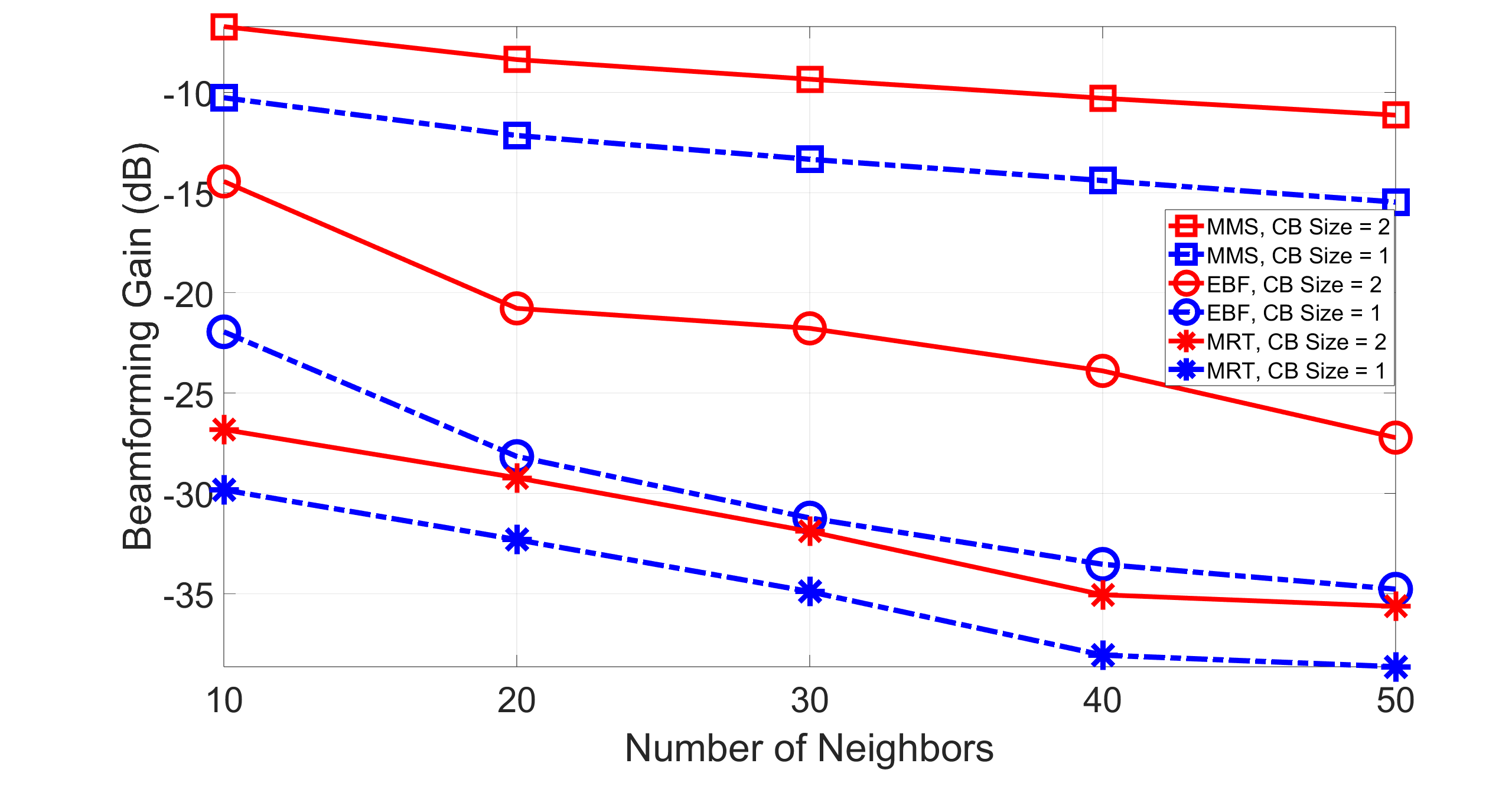}
	\caption{Performance evaluation of the beamforming gain as a function of the number of neighbors in the initial neighborhood list for a circular trajectory with the neighborhood parameter set to $5$.}
	\label{fig:CircPathComparison_Ndata1000_10LocalNbrs}
	\vspace{-0.3cm}
\end{figure}

Figure~\ref{fig:CircPathComparison_Ndata1000_10LocalNbrs} shows the beamforming gain (dB) as a function of the size of the initial neighborhood list for the codebook sizes of $1$ and $2$. We clearly see that the MMS algorithm outperforms both the EBF and the \gls{mrt} by a large margin for both the codebook sizes of $1$ and $2$. We also observe that, as the number of neighbors increases, the beamforming gain decreases for all the schemes. This can be proved mathematically: Suppose the size of the neighborhood list is denoted by $K$. For simplicity, let the size of the codebook be $1$. Then,
\begin{align}
	\min_{k\in [K]}\frac{\arrowvert\mathbf{h}_k^H \mathbf{f}\arrowvert^2}{\Arrowvert \mathbf{h}_k\Arrowvert^2}&\le\min_{k\in [K-1]}\frac{\arrowvert\mathbf{h}_k^H \mathbf{f}\arrowvert^2}{\Arrowvert \mathbf{h}_k\Arrowvert^2}\label{eqn:SimResults1}\\
	\implies \max_{\mathbf{f}}\min_{k\in [K]}\frac{\arrowvert\mathbf{h}_k^H \mathbf{f}\arrowvert^2}{\Arrowvert \mathbf{h}_k\Arrowvert^2}&\le\max_{\mathbf{f}}\min_{k\in [K-1]}\frac{\arrowvert\mathbf{h}_k^H \mathbf{f}\arrowvert^2}{\Arrowvert \mathbf{h}_k\Arrowvert^2}.\label{eqn:SimResults2}
\end{align}
As mentioned in \eqref{eqn:SimResults1}, the minimum beamforming gain cannot increase when we add more vectors in the neighborhood list. Therefore, the max-min-sum beamforming gain achieved by any algorithm does not improve as the size of the neighborhood list increases. Also, we observed in our simulations that, when $K$ is set to $1$ (i.e., the current \gls{csi} is exact), all the algorithms converge to the \gls{mrt} beamformer.

\begin{figure}[!t]
	\centering
	\includegraphics[width=\columnwidth]{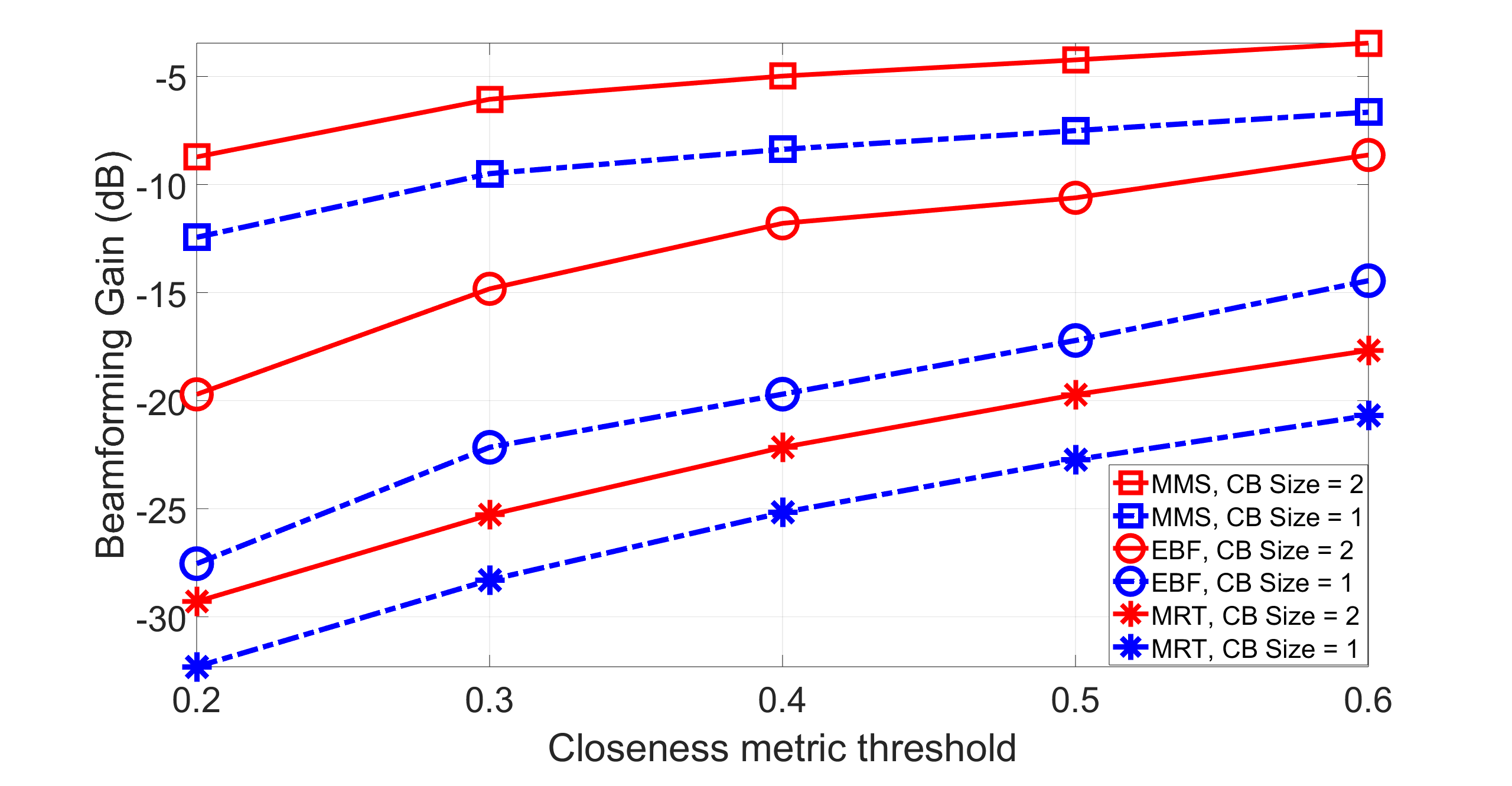}
	\caption{Performance evaluation of the beamforming gain as a function of the closeness-metric threshold for a circular trajectory with the neighborhood parameter set to $5$.}
	\label{fig:CircPathComparison_Ndata1000_10LocalNbrs_ClosenessMetricVsBFGain}
	\vspace{-0.3cm}
\end{figure}

Figure~\ref{fig:CircPathComparison_Ndata1000_10LocalNbrs_ClosenessMetricVsBFGain} shows the beamforming gain (dB) as a function of the closeness-metric threshold for the codebook sizes of $1$ and $2$. The closeness-metric threshold is used to determine the elements of the initial neighborhood channel list. We compute the closeness-metric between the current outdated \gls{csi} and all the entries in the channel database, and pick the elements whose values are above the threshold into the initial neighborhood channel list. Then, we obtain the total neighborhood list as mentioned in the neighborhood generation procedure. {We vary the closeness-metric threshold from $0.2$ to $0.6$ to obtain Figure~\ref{fig:CircPathComparison_Ndata1000_10LocalNbrs_ClosenessMetricVsBFGain}.} As the closeness-metric threshold increases, the size of the initial neighborhood channel list decreases, which translates to an increase in the beamforming gain as proved in \eqref{eqn:SimResults2}. We also see that the MMS unanimously outperforms both the EBF and \gls{mrt} by a large margin.

\begin{figure}[!t]
	\centering
	\includegraphics[width=\columnwidth]{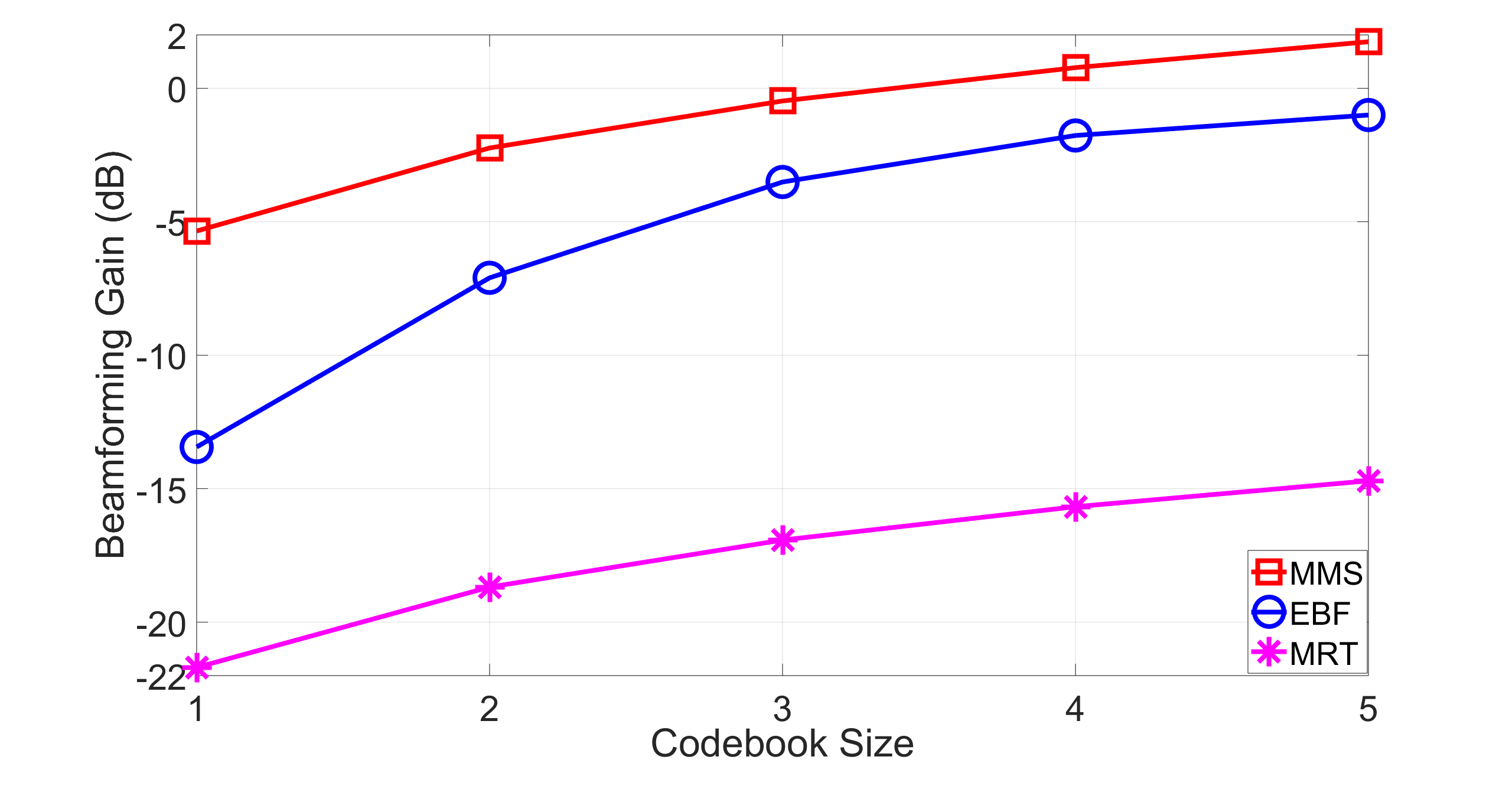}
	\caption{Performance evaluation of the beamforming gain as a function of the codebook size for a zig-zag trajectory with the neighborhood parameter set to $5$. The closeness-metric threshold is set to $0.766$ (which corresponds to a maximum angle separation of $40$~degrees between the current outdated \gls{csi} and the elements in the database).}
	\label{fig:SinePath_Ndata2000_10LocalNbrs_CBSzvsBFGain_AngleThreshold40degrees}
	\vspace{-0.65cm}
\end{figure}

Figure~\ref{fig:SinePath_Ndata2000_10LocalNbrs_CBSzvsBFGain_AngleThreshold40degrees} shows the beamforming gain (dB) as a function of the codebook size when the closeness-metric threshold is set to $0.766$ (which corresponds to a maximum angle separation of $40$~degrees between the current outdated \gls{csi} and the elements in the database). We again see that the MMS clearly achieves a much larger beamforming gain than the EBF and \gls{mrt} schemes. We also observe that the beamforming gain achieved by the EBF with a codebook size of $4$ can be achieved with only $2$ beamforming vectors by the developed MMS procedure. Even with a codebook size of $5$, the \gls{mrt} cannot achieve a beamforming gain of the MMS scheme with only one beamforming vector in the codebook.

\section{Conclusions}
In this paper, we have developed a novel data-driven mechanism to select a channel neighborhood from a database and a principled approach to design a beamforming codebook to maximize the minimum sum beamforming gain across all the channels in the neighborhood list. Our approach is a proof of concept that customized solutions based on environment specific scenarios are indeed important to address the requirements of future wireless applications. Further, we plan to research on designing more advanced data-driven methods for multi-antenna \glspl{ue} and multi-user massive \gls{mimo}.
\bibliographystyle{IEEEtran}
\bibliography{IEEEabrv,References}

\begin{thebibliography}{10}
\providecommand{\url}[1]{#1}
\csname url@samestyle\endcsname
\providecommand{\newblock}{\relax}
\providecommand{\bibinfo}[2]{#2}
\providecommand{\BIBentrySTDinterwordspacing}{\spaceskip=0pt\relax}
\providecommand{\BIBentryALTinterwordstretchfactor}{4}
\providecommand{\BIBentryALTinterwordspacing}{\spaceskip=\fontdimen2\font plus
\BIBentryALTinterwordstretchfactor\fontdimen3\font minus
  \fontdimen4\font\relax}
\providecommand{\BIBforeignlanguage}[2]{{%
\expandafter\ifx\csname l@#1\endcsname\relax
\typeout{** WARNING: IEEEtran.bst: No hyphenation pattern has been}%
\typeout{** loaded for the language `#1'. Using the pattern for}%
\typeout{** the default language instead.}%
\else
\language=\csname l@#1\endcsname
\fi
#2}}
\providecommand{\BIBdecl}{\relax}
\BIBdecl

\bibitem{Ahmet_Globecom_2022}
A.~Kaplan, J.~Vieira, and E.~G. Larsson, ``Dynamic range improvement in
  bistatic backscatter communication using distributed {MIMO},'' in
  \emph{Proc.\ Globecom}, 2022, pp. 2486--2492.

\bibitem{Perera_IEEECOMTUTs_2018}
T.~D. Ponnimbaduge~Perera, D.~N.~K. Jayakody, S.~K. Sharma, S.~Chatzinotas, and
  J.~Li, ``Simultaneous wireless information and power transfer ({SWIPT}):
  Recent advances and future challenges,'' \emph{{IEEE} Trans. Commun. Surveys
  Tuts.}, vol.~20, no.~1, pp. 264--302, 2018.

\bibitem{Giordani_COMMSURVs_2019}
M.~Giordani, M.~Polese, A.~Roy, D.~Castor, and M.~Zorzi, ``A tutorial on beam
  management for {3GPP NR} at mmwave frequencies,'' \emph{{IEEE} Trans. Commun.
  Surveys Tuts.}, vol.~21, no.~1, pp. 173--196, 2019.

\bibitem{3GPP_TR38843_2017NR}
``{3GPP TR} 38.802 {V}14.2.0 (2017-09); study on new radio access technology
  physical layer aspects ({R}elease 14),'' 2019.

\bibitem{3GPP_TR38843_2022NR}
``{3GPP TR} 38.843 {V}0.0.0 (2022-05); study on artificial intelligence
  ({AI}/machine learning ({ML}) for {NR} air interface ({R}elease 18),'' 2022.

\bibitem{Mustafa_VTCS_2022}
M.~F. Özkoç, C.~Tunc, and S.~S. Panwar, ``Data-driven beamforming codebook
  design to improve coverage in millimeter wave networks,'' in \emph{Proc.\
  {VTC} (Spring)}, 2022, pp. 1--7.

\bibitem{Karipidis_TSP_2008}
E.~Karipidis, N.~D. Sidiropoulos, and Z.-Q. Luo, ``Quality of service and
  max-min fair transmit beamforming to multiple cochannel multicast groups,''
  \emph{{IEEE} Trans. Signal Process.}, vol.~56, no.~3, pp. 1268--1279, 2008.

\bibitem{Christopoulos_SPAWC_2015}
D.~Christopoulos, S.~Chatzinotas, and B.~Ottersten, ``Multicast multigroup
  beamforming for per-antenna power constrained large-scale arrays,'' in
  \emph{Proc.\ SPAWC}, Jun. 2015, pp. 271--275.

\bibitem{MinDong_TSP_2020}
M.~Dong and Q.~Wang, ``Multi-group multicast beamforming: Optimal structure and
  efficient algorithms,'' \emph{{IEEE} Trans. Signal Process.}, vol.~68, pp.
  3738--3753, 2020.

\bibitem{cvxr}
M.~Grant and S.~Boyd, ``{CVX}: Matlab software for disciplined convex
  programming, version 2.1,'' \url{http://cvxr.com/cvx}, Mar. 2014.

\bibitem{gb08}
------, ``Graph implementations for nonsmooth convex programs,'' in
  \emph{Recent Advances in Learning and Control}, ser. Lecture Notes in Control
  and Information Sciences, V.~Blondel, S.~Boyd, and H.~Kimura, Eds.\hskip 1em
  plus 0.5em minus 0.4em\relax Springer-Verlag Limited, 2008, pp. 95--110.

\bibitem{marks1978general}
B.~R. Marks and G.~P. Wright, ``A general inner approximation algorithm for
  nonconvex mathematical programs,'' \emph{Oper. Res.}, vol.~26, no.~4, pp.
  681--683, 1978.

\bibitem{Hunter_AmerStat_2004}
D.~R. Hunter and K.~Lange, ``A tutorial on {MM} algorithms,'' \emph{The
  American Statistician}, vol.~58, no.~1, pp. 30--37, 2004.

\bibitem{Wilding_Sensors_2018}
T.~Wilding, S.~Grebien, U.~M{\"u}hlmann, and K.~Witrisal, ``Accuracy bounds for
  array-based positioning in dense multipath channels,'' \emph{Sensors},
  vol.~18, no.~12, p. 4249, 2018.

\bibitem{Deutschmann_ICCWS_2022}
B.~J.~B. Deutschmann, T.~Wilding, E.~G. Larsson, and K.~Witrisal,
  ``Location-based initial access for wireless power transfer with physically
  large arrays,'' in \emph{Proc.\ {ICC Workshops}}, 2022, pp. 127--132.

\bibitem{Leitinger_JSAC_2015}
E.~Leitinger, P.~Meissner, C.~Rüdisser, G.~Dumphart, and K.~Witrisal,
  ``Evaluation of position-related information in multipath components for
  indoor positioning,'' \emph{{IEEE} J. Sel. Areas Commun.}, vol.~33, no.~11,
  pp. 2313--2328, 2015.

\end{thebibliography}
\end{document}